# Quasi-coherent thermal emitter based on refractory plasmonic materials

Jingjing Liu[1], Urcan Guler[2], Alexei Lagutchev[1], Alexander Kildishev[1], Oana Malis[3], Alexandra Boltasseva[1,4], Vladimir M. Shalaev[1,*]

[1]*School of Electrical and Computer Engineering and Birck Nanotechnology Center, Purdue University, West Lafayette, IN 47907, USA*

[2]*Nano-Meta Technologies, Inc., 1281 Win Hentschel Boulevard, West Lafayette, IN 47906, USA*

[3]*Department of Physics, Purdue University, West Lafayette, IN 47907, USA*

[4]*DTU Fotonik, Department of Photonics Engineering, Technical University of Denmark, Lyngby, DK-2800, Denmark*

[*]*shalaev@purdue.edu*

**Abstract**: The thermal emission of refractory plasmonic metamaterial – a titanium nitride 1D grating – is studied at high operating temperature (540 °C). By choosing a refractory material, we fabricate thermal gratings with high brightness that are emitting mid-infrared radiation centered around 3 μm. We demonstrate experimentally that the thermal excitation of plasmon-polariton on the surface of the grating produces a well-collimated beam with a spatial coherence length of 32λ (angular divergence of 1.8°) which is quasi-monochromatic with a full width at half maximum of 70 nm. These experimental results show good agreement with a numerical model based on a two-dimensional full-wave analysis in frequency domain.

## 1. Introduction

Despite recent progress in developing coherent sources of radiation in the mid-infrared, thermal emitters enhanced with modern nanoplasmonic technology may appear quite competitive for applications such as sensing, spectroscopy and local heat transfer [1]. With proper modifications of narrowing the linewidth and improving directionality of emission, it is possible to design a low-cost, scalable, and ultra-compact mid-IR emitter using quasi-coherent thermal radiation. One of the promising ways to achieve active control over direction and spectral properties of thermal emission is to employ surface structure supporting electromagnetic eigenmodes [2-6].

To date various quasi-coherent thermal sources have been studied. Researchers have made narrow-band thermal emitters from metallic photonic crystals [7, 8], silicon photonic crystals coupled with quantum wells [9, 10], arrays of deep metal grooves [11] and micropatches [12]. Guided modes [13] and coupled resonant cavities [3] have been also proposed to generate spatial coherence.

Thermal excitation of delocalized modes is one notable way to obtain spatial coherence. In this case, electrons or phonons are no longer uncorrelated but semi-coherently oscillating along the decay length of the mode propagation. Hence their emitted fields are spatially coherent to a degree and, if properly coupled to the far field, can produce constructive interference along certain direction in space. One practical way to achieve this is to use grating structures. There have been numerous studies about coupling surface waves to free-space waves using 1D gratings, e.g., [2, 4, 14]. Most of the articles deal with the thermal emission of gratings made of polar materials and noble metals. In polar material, the enabling mechanism is coupling between the optical phonons and light. However these surface phonon-polaritons exist only in the Reststrahlen band and their emission wavelength can be tuned only over a relatively narrow range [15]. On the other hand, metals are well known to support interface modes called surface plasmon-polaritons in a rather broad wavelength range. However, noble metals have thermal stability limits, so heating a metallic structure eventually leads to delamination, diffusion with aggregation into islands and eventually melting and evaporation [16, 17]. According to Stefan-Boltzmann's law, intensity of thermal emission source is proportional to the fourth power of the temperature ($\sim T^4$). It is therefore extremely beneficial to replace metals with materials having similar plasmonic properties, albeit capable of withstanding markedly higher temperatures in order to increase the thermal source brightness and further add the benefit of blue-shifting the emission wavelength peak to near-infrared. Recently, transition metal nitrides have been proposed as refractory plasmonic materials that can stand the harsh environments imposed by high temperature applications [18, 19]. In this letter, we demonstrate coherent thermal emission from a refractory plasmonic structure, titanium nitride (TiN) grating, at the wavelength around 3 μm. Generally speaking, the emission characteristics of TiN grating can be tuned to cover all the near and mid-infrared wavelengths by altering the geometry of the surface. However, we have chosen to demonstrate a source with the emission wavelength close to 3 μm as the vibrational C-H stretch resonances present in vast majority of organic molecules are located close to this wavelength. This makes 3 μm range specifically attractive for sensing applications. Furthermore, durability of TiN nanostructures has been demonstrated at 800°C [17] while similar gold nanostructures was previously reported to work around 400°C [20]. Due to elevated working temperatures and $T^4$ dependence, the radiation intensity of TiN nanostructures can be approximately six times higher compared to similar gold nanostructures. This superior thermal stability makes TiN an excellent candidate material for substantial improvement on emission intensity of coherent thermal sources vs. conventional noble metals.

## 2. Numerical calculations of the emissivity

Kirchhoff's law states that the angular and frequency dependent emissivity of a reciprocal object is equal to its corresponding absorptivity. For an opaque sample, transmission is negligible. Kirchhoff's law implies that the emissivity is equal to $1-R$, where $R$ is the reflectance of the sample. Here, the reflectance of the grating was calculated based on the two dimensional spatial harmonic analysis (SHA) [21], also known as Fourier modal method (FMM) or rigorous coupled-wave analysis (RCWA). 2D-SHA solves Maxwell's equations for periodic grating structures, extracts the magnetic field as the eigenvectors for *p*-polarization, cascades the magnetic fields for multiple layers using the Thomas algorithm for structured transfer matrices, and finally gives the reflectance. The permittivity of TiN at room temperature was measured using variable-angle spectroscopic ellipsometer by fitting the experimental data with the Drude-Lorentz model.

Fig. 1(b) shows the calculated emissivity of the grating along surface normal for *p*- polarization (*E*-field is in the incidence plane), as only *p*-polarized waves are coupled to surface plasmons in 1D metal gratings in the direction perpendicular to gratings lines. The TiN grating with a period Λ of 3 µm, a groove height h of 100 nm, and a filling factor (q = w/ Λ) of 0.537 was modelled with 40-nm SiN cover layer on top, which was used in experiment to prevent oxidation of TiN at elevated temperatures. The schematic of the sample is shown in Fig. 1(a). Fig. 1(c) indicates that at the wavelength of 3.05 µm, the calculated directionality of the *p*-polarized emission has a divergence angle of mere 0.7°. This is a signature of the spatial coherence of the source. The TiN grating structure theoretically yields a thermal emission with a spatial coherence length $l_c = \lambda / \Delta\theta = 82\lambda$ and a quality factor $Q = \lambda / \Delta\lambda = 87$ ($\Delta\lambda = 35$ nm). For comparison, we compute the gold structures under the same geometry with permittivity approximated by Johnson-Christy model with a loss factor of three [22]. We obtain coherence length $l_c = 204\lambda$ (angular divergence is 0.28°) and quality factor $Q = 231$ ($\Delta\lambda = 13$ nm) from gold grating sturctures at the wavelength of 3 µm. The difference is mainly due to the fact that TiN is less metallic than gold, so the electromagnetic field penetrates more into the TiN. It follows that in TiN the propagating length of the surface wave is smaller than that of gold. However for practical purposes the benefit of higher radiation intensity of TiN grating is likely to outweigh its somewhat inferior coherence characteristics as compared to identical gold structure.

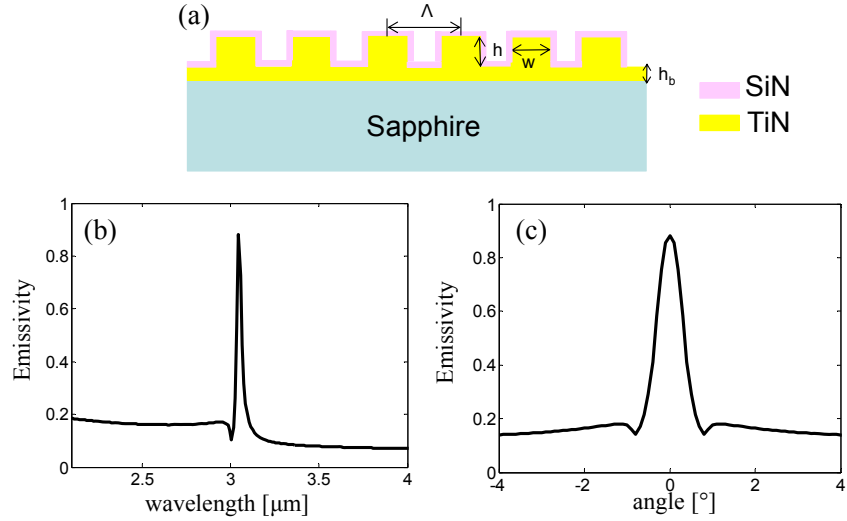

Fig.1. (a) Schematics of the TiN grating structure. Calculated emissivity of a TiN grating along surface normal as a function of (b) wavelength and (c) angle. The parameters of the grating are: period, $\Lambda = 3$ μm, filling factor, $q = w/\Lambda = 0.537$, height, $h = 100$ nm, and height of bottom TiN layer $h_b = 50$ nm.

## 3. Sample fabrication and emissivity measurements

Fig. 2(a) shows the scanning electron microscope image of the TiN grating. A top-down approach is used for sample fabrication. First, 150 nm TiN is grown on a c-Sapphire substrate (<111> orientation) in a nitride sputtering system. Then, grating patterns were defined on the negative tone resist, hydrogen silsesquioxane (HSQ) using standard electron beam lithography (EBL). The lithography process began by spin–coating HSQ on the TiN film and subsequent soft–baking. A 1.6 mm × 1.6 mm grating lines were patterned by exposing the resist, and the resist was developed in tetramethylammonium hydroxide (TMAH) at room temperature. The pattern was transferred into the TiN layer by dry etching in an inductively coupled plasma (ICP) reactive ion etching system using $Cl_2$. Finally, HSQ mask was rinsed away by hydrofluoric acid (HF). Subsequently a 40nm thick anti-oxidation layer of SiN was grown on top of TiN grating by chemical vapor deposition.

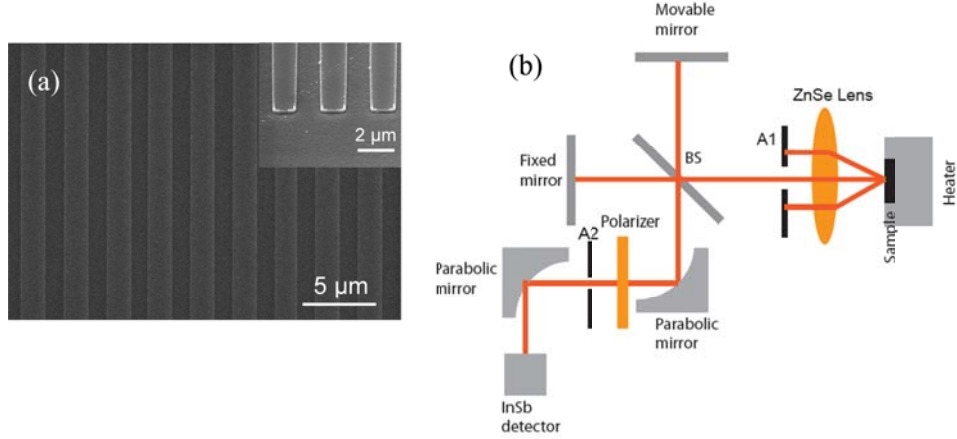

Fig.2. (a) SEM image of the TiN grating, (b) Schematic of the experimental setup for thermal emission measurements of TiN grating. BS: KBr beamsplitter, A1, A2: iris apertures.

Prepared sample was electrically heated to 540°C in air by commercially available heating stage (HCP621, Instec, Inc.). Sample emission was directed into an external port of FTIR spectrograph (Nexus 870, Thermo Fisher Scientific, Inc.) equipped with $CaF_2$ coated beamsplitter and liquid nitrogen-cooled InSb detector. It should be noted that the temperature for the experiment was limited by the heating stage in use. Thus our potential emissivity gain by raising the temperature from 400°C to 540°C was still respectable 2-fold. As we have experimentally shown before, the sample could be heated up to 800°C without structure degradation [17] for the full 6-fold emissivity increase. The experimental setup is shown in Fig. 2(b). The emission was collected by a ZnSe lens (f = 63.5 mm) and collimated into the input port of FTIR. By varying the position of the aperture A1, we analyze the portion of light emitted at a range of angles defined by the aperture diameter. In the setup, aperture diameter was 1.5 mm, with corresponding angular resolution $\alpha = 1.35°$. The aperture A1 had to be kept large enough to get sufficient signal, thus the angular resolution was limited by FTIR signal to noise ratio. Another aperture A2 was positioned in the conjugate plane of the sample. The size of this aperture defined the collection area on the sample. The measurements were performed with a spectral resolution of 2 $cm^{-1}$ corresponding to 2 nm at $\lambda = 3\mu m$. Each spectrum was averaged over 256 scans. Carbon black was used as the blackbody reference to extract the spectral emissivity of the sample. The emission spectrum was collected at 300°C since the carbon black sample evaporates at temperatures beyond this point. The emission spectrum at higher temperature could be estimated by [23]

$$S_B(\lambda, T_2) - S_R(\lambda, T_R) = (S_B(\lambda, T_1) - S_R(\lambda, T_R))\frac{P(\lambda, T_2) - P(\lambda, T_R)}{P(\lambda, T_1) - P(\lambda, T_R)} \qquad (1)$$

Here, $S_B(\lambda,T_2)$ is the estimated spectrum of the blackbody radiation at temperature $T_2 = 540°C$, and $S_B(\lambda,T_1)$ is the experimentally measured spectrum of the blackbody radiation at temperature $T_1 = 300°C$. $S_R(\lambda,T_R)$ is the spectrum of the room radiation. $P(\lambda,T)$ is the Planck radiation function at temperature $T$. The emission spectra of the sample are measured at the temperature of 540°C. The emissivity is calculated following the approach used in reference [23]:

$$\varepsilon(\lambda,T) = \frac{S_S(\lambda,T) - S_R(\lambda,T_R)}{S_B(\lambda,T) - S_R(\lambda,T_R)} \qquad (2)$$

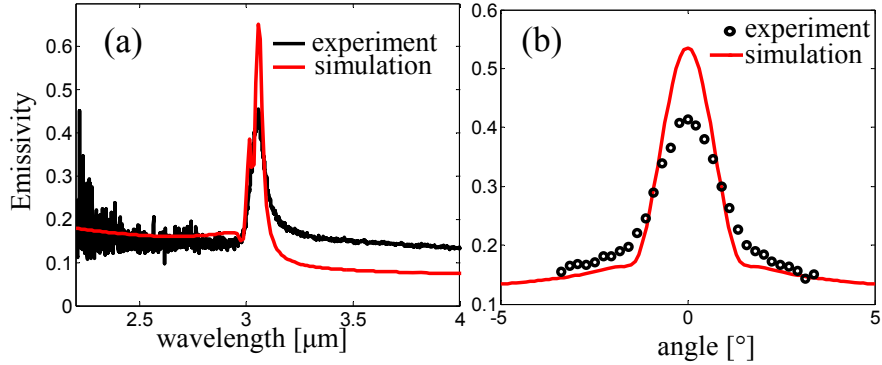

Fig.3 (a) Measured and calculated emissivities as a function of the wavelength for emission angle θ = 0° at 540 °C. (b) Angular profile of emissivity at the wavelength of 3.06 μm vs. calculation.

Fig. 3(a) shows the measured emissivity in the direction normal to the surface. Theoretically the grating with a pitch size of 3 μm is shown to have a high emissivity at 3.05 μm. Experimentally we measured a narrow peak at the wavelength of 3.06 μm with full width at half maximum equal to 70 nm corresponding to a Q factor of 43. The measured wavelength profile is the integrated emission from the TiN grating over the detection angle of 1.35° normalized by the emission from blackbody reference. The simulation followed the experimental setup with averaging of the emissivity over the same angular range. Thus TiN grating appears to yield a narrow-band emission with a spatial coherence length $l_c = \lambda / \Delta\theta = 32\lambda$ (angular divergence of 1.8°) as shown in Fig. 3(b). We also measured the emissivity of planar unpatterned region to be 0.12 which is slightly higher than the theoretical calculation 0.08. The slight deviation is attributed to surface roughness and structure imperfections inherent in nanofabricated samples.

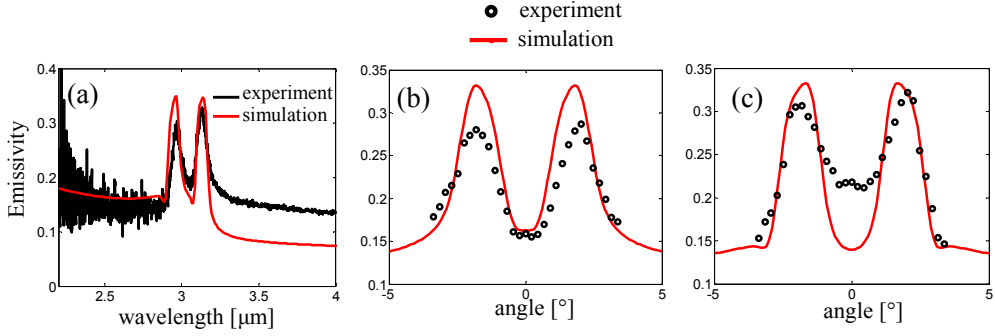

Fig.4 (a) Measured vs. calculated emissivity as a function of the wavelength for θ = 1.8° at 540 °C. (b) Angular profile of the emissivity at the wavelength of (b) 2.96 μm and (c) 3.13 μm.

The direction of the measured emission can be chosen by moving the aperture A1 off the optical axis of the ZnSe lens According to grating dispersion equation $\frac{2\pi}{a} \pm \frac{2\pi}{\lambda}\sin\theta = \frac{2\pi}{\lambda}\sqrt{\frac{\varepsilon_m \varepsilon_d}{\varepsilon_m + \varepsilon_d}}$ [24] ($a$ is the grating period, $\varepsilon_m$ and $\varepsilon_d$ are permittivities of metal and dielectric), two wavelength peaks should be present at each emission direction. The peaks corresponding to the θ = 1.8° direction are shown in Fig.4 (a). Also shown on Fig.4 (b, c) are the measured and calculated angular distribution of emissivities at the wavelengths of 2.96 μm and 3.13 μm, respectively. As seen from the graphs, experimental angular dependencies of emission follow the computational model reasonably well with calculated angular maxima deviating from experimentally measured values by about 0.2°.

In Fig. 5 we present the full diagram of emission spectral peak positions versus observation angles. The spectral positions of the peaks are well matched with calculated dispersion plot of TiN grating. As it was noted in earlier reports [4], the model of grating structure under investigation predicts the existence of a small ($\delta\omega / \omega_0 = 3.3 \times 10^{-3}$) bandgap in the center of calculated dispersion plot as seen in the inset of Fig.5. The presence of such bandgap feature results in a significant increase the local density of states of surface plasmon-polaritons which is another way to describe observed emission enhancement.

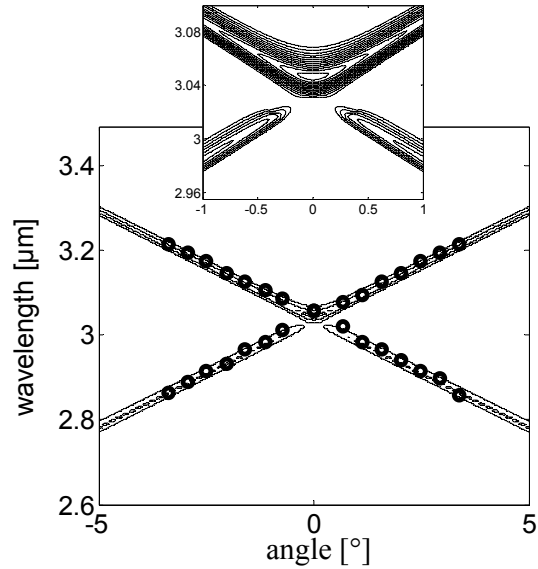

Fig.5 Experimental (black dots) and calculated emissivity (contours) spectrum vs. emission angles. Expanded central area of the plot shown in the inset.

## 4. Conclusions

To conclude, we have demonstrated a quasi-monochromatic and spatially coherent thermal emission from a grating made of refractory plasmonic material – TiN. The substantial improvement of the coherent properties of the thermally excited electromagnetic field is demonstrated to be related to the excitation of surface plasmon-polaritons. By engineering the infrared grating sources according to the developed computational model, we have realized a directional and narrow-band thermal source of 3 μm light. As the limited brightness of the infrared source presents a challenge for many applications, by switching to the refractory material for the grating-based source and raising the working temperature from 400°C to 540°C, one can achieve potentially 2-fold (6-fold at 800°C) increase in radiation power compared to similar structure made of gold. This presents a significant benefit for the cases when low-cost and compact, yet narrowband and low-divergence sources in the mid-infrared are required, such as in various sensing and spectrometry applications.

**Acknowledgements**

This work was partially supported by NSF Center for Photonic and Multiscale Nanomaterials, ARO MURI Grant 56154-PH-MUR (W911NF-09-1-0539).